\documentclass[pre,aps,showpacs]{revtex4}
\usepackage{amsmath,amsfonts,bm}
\usepackage[dvips]{graphicx}
\usepackage[floatfix]{epsfig}
\usepackage{subfigure}

\begin{document}

\renewcommand{\thesubfigure}{(\alph{subfigure})}

\title{Numerical study of polymer tumbling in linear shear flows}
\author{A. Puliafito$^{1,3}$, K. Turitsyn$^{1,2,3}$}

\affiliation{
$^1$ INLN, UMR 6618 CNRS, 1361, Route des Lucioles, 06560 Valbonne, France\\
$^2$ Landau Institute for Theoretical Physics, Kosygina 2, Moscow 119334, Russia \\
$^3$ Theoretical Division, Los Alamos National Laboratory, Los Alamos, NM 87545, USA
}
\date{\today}

\begin{abstract}
We investigate numerically the dynamics of a single polymer in a linear shear flow.
The effects of thermal fluctuations and randomly fluctuating velocity gradients
are both analyzed.
Angular, elongation and tumbling time statistics are measured numerically.
We perform analytical calculations and
numerical simulations for a linear single-dumbbell polymer model
comparing the results with previous theoretical and experimental studies.
For thermally driven polymers the balance between relaxation and
thermal fluctuations plays a fundamental role, whereas for random velocity
gradients the ratio between the intensity of the random part and the
mean shear is the most relevant quantity.
In the low-noise limit,
many universal aspects of the motion of a polymer
in a shear flow can be understood in this simplified framework.
\end{abstract} \maketitle

\section{Introduction}
Thanks to recent improvements in experimental techniques it is nowadays possible to 
follow the motion of individual molecules in 
solvents~\cite{97PSC,98SC,99SBC,00HSL,01HSBSC,04GCS,05GS,chu2,05STSC}. 
The characterization of polymer dynamics at the level of a single molecule is a first 
step towards the understanding of mechanical interactions between biomolecules 
(see e.g.~\cite{ladoux,chu,manneville,larson,cui,hegner,yin,wuite}), of the fundamental 
rheology of polymer solutions, and of the viscoelastic properties of more complex flows 
(see for example~\cite{04GS} and references therein for elastic turbulence). \\
Recently measurements of elongation and orientation with respect to simple external 
flows have been performed~\cite{05GS,05STSC} in order to analyze how the conformation
of a single molecule can be modified by an external field.\\
As the number of degrees of freedom needed to fully describe a macromolecule is 
extremely high it is necessary to formulate theoretical models, to verify them with 
numerical simulations, to simplify the problem , and to understand which observables play
a key role. \\
Large numbers of papers have been written on the subject of single polymer
dynamics in shear flows from experimental, numerical and theoretical viewpoint by Chu,
Larson, Shaqfeh and their respective collaborators.\\
A polymer molecule in a plane, linear, steady shear flow is oriented in the flow 
direction by the velocity field. As the rotational motion of the polymer is determined 
only by the velocity difference in the space points, when it is aligned along the 
shear direction the external flow effect becomes negligible, and the thermal noise is the
most relevant external force. Thermal fluctuations can skew the polymer to regions where
the external flow is relevant again. In these cases the shear flow can induce a fast 
rotation and align the polymer again along the (reverse) flow direction, i.e. the polymer
tumbles~\cite{89Liu,99SBC,00HSL,chu2,05STSC,05GS}. \\
This phenomenon can happen via several conformational pathways due to the complexity
of the motion of a polymer molecule (see for example \cite{chu2,05STSC}), and can be
fully described only by taking into account all the degrees of freedom of the polymer.
Unfortunately in the framework of these complex polymer models it is very difficult to
obtain analytical results.\\
The main goal of our paper is to test numerically the predictions of
recent theoretical and experimental studies~\cite{04CKLTa,05GS} in a framework in which 
analytical results can be obtained~\cite{T04}, and to analyze the statistics of the
tumbling times, i.e. the time between two subsequent flips of the
polymer~\cite{89Liu,99SBC,05GS}. The simplest model that reproduces
qualitatively the behavior of polymers is the dumbbell
model~\cite{book}. This model allows one to carry out some analytical
calculations in the case which we analyze~\cite{72HL,04CKLTa,04CKLTb,T04},
it is very easy and fast to simulate numerically (see
sec.~\ref{sec:numerics} and~\cite{oett}) and reproduces qualitatively recent
experimental results~\cite{99SBC,05GS}. \\
The paper is organized as follows: in sec.~\ref{sec:numerics} the evolution
equation of the polymer and the numerical methods are briefly explained.
Sec.~\ref{sec:therm} is devoted to the analysis of the statistics of
thermal fluctuations of a flexible polymer placed into a linear
shear flow. In our work we present the analysis of the stationary
distribution of the polymer end-to-end vector and we study the
distribution function of the polymer tumbling time, which can be
measured experimentally. In sec.~\ref{sec:rigid} we study the
angular dynamics of strongly elongated polymers, for which the size
fluctuations are negligible. Finally in sec.~\ref{sec:cramer} we
study the elongation statistical properties of the end-to-end vector
in a random flow with a large mean shear.

\section{Basic relations and numerical analysis}\label{sec:numerics}
We wish to analyze the behavior of a polymer in a generic
simple shear flow experiencing the Langevin force~\cite{72HL}. In
general there are two effects of the velocity field on a polymer
molecule: the Lagrangian advection of the polymer and the
elongation/relaxation dynamics due to velocity gradients. In all the
cases discussed in this paper we disregard the Lagrangian dynamics by
staying in the reference frame of the polymer center of mass. For
the internal degrees of freedom of the polymer we use the simple
dumbbell model~\cite{book}, leaving the analysis of more realistic
models for future studies~\cite{future}. In this case the basic
equation describing the evolution of the polymer end-to-end vector
$\bm R $ has the following form:
\begin{equation} \label{eq:main}
 \dot{R}_i  = \sigma_{ij} R_j - \gamma R_i + \sqrt{ \frac{2 \gamma R_0^2}{3}}\xi_i ,
\end{equation}
where $\sigma_{ij} =\partial_j v_i$ is the velocity gradient matrix, $\gamma$ is
the polymer relaxation rate, $\bm \xi$ is the thermal noise term,
which has white-noise statistics: $\langle
\xi_i(t)\xi_j(t')\rangle=\delta_{ij}\delta(t-t')$, and $R_0$ is the equilibrium length in
the absence of an external field. \\
For incompressible flows there is no ambiguity in the discretization of the stochastic
differential equation \eqref{eq:main},
and no additional contact term must be taken into account, so we resort to the
Euler-It\^o scheme without any loss of generality.\\
When the gradient $\sigma$ contains only a steady linear shear contribution we can
write down the formal solution of \eqref{eq:main} as:
\begin{gather}\label{eq:num2}
 \dot{R}_i(t)  = R_i(0)e^{-\gamma t}
+e^{-\gamma t}\int_0^t e^{\gamma t'} s_{ij} R_j(t')\delta_{ix}\delta_{jy} dt'+\notag\\
-\gamma e^{-\gamma t} \int_0^t e^{\gamma t'} R_i(t') dt'
+ \sqrt{\frac{2 \gamma R_0^2}{3}}e^{-\gamma t}\int_0^t e^{\gamma t'}\xi_i(t') dt'.
\end{gather}
This equation can be discretized, and the terms containing the thermal noise can be
rewritten as new gaussian variables with amplitudes that can be computed directly, so
that the final solution reads:
\begin{gather}
R_x^{k+1}=R_x^{k}e^{-\gamma \Delta t}+\sqrt{\frac{R_0^2}{3}
(1-e^{-2\gamma \Delta t})}\eta_x^{k}+\notag 
+s\Delta t e^{-\gamma \Delta t}R_y^{k} +s\frac{R_0}{\sqrt{3}}\eta_y^{k} \times \notag\\\times
\sqrt{\left[\frac{1}{2\gamma^2}(1-e^{-2\gamma \Delta t})
-\frac{1}{\gamma}\Delta t e^{-2\gamma \Delta t}-\Delta t^2 e^{-2\gamma \Delta t} \right]}
\\
  R_i^{k+1}=R_i^{k}e^{-\gamma \Delta t}+\sqrt{\frac{R_0^2}{3}
(1-e^{-2\gamma \Delta t})}\eta_i^{k}\, ,\,i=y,z
\end{gather}
where $\langle \eta_i^{k}\eta_j^{l}\rangle=\delta_{ij}\delta^{kl}$. The subscripts stand
for the cartesian coordinates while the superscripts stand for the discretized time.
As expected in the limit $\Delta t \rightarrow 0$ there is no contribution of $\eta_y$
in the first equation and the amplitudes are the same as in eq. \eqref{eq:main}.\\
In the case of a random velocity gradient, $w_{ij}$, we have to generate the variables
$w_{ij}^{k}$ such that they have the prescribed correlation function:
\begin{equation}\label{eq:corr}
 \langle \sigma_{ij}(t)\sigma_{kl}(t')\rangle= D
 \delta(t-t')(4\delta_{ik}\delta_{jl} -\delta_{il}\delta_{kj}-\delta_{ij}\delta_{kl})\, ,
\end{equation}
where again the Dirac delta function is substituted by a Kronecker symbol.
In this case the modulus of the vector $\bm R$ grows indefinitely so that we can
normalize it at each time step. In order to describe the elongation properties we can
compute the maximum Lyapunov exponent $\lambda$ and the corresponding finite time
Lyapunov exponent $\lambda_T$~\cite{80BGGS}.

\section{Thermally driven polymers} \label{sec:therm}
\begin{figure}[t]
\includegraphics[width=0.42\textwidth]{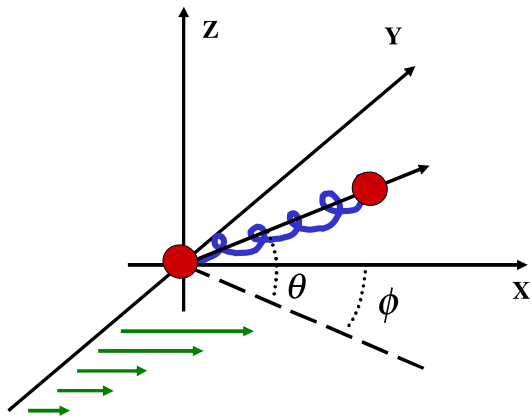}
 \caption{Schematic figure explaining polymer orientation geometry.}
 \label{fig:orient}
\end{figure}
In this section we will examine the case of a linear steady shear flow in the
plane $XY$: $\sigma_{ij} = s\delta_{ix}\delta_{jy}$, where $s$ is the shear rate.
Eq. \eqref{eq:main} has an explicit solution of the following form:
\begin{gather}
 R_i(t) = \exp(-\gamma t)W_{ij}(t)R_j(0) +\int_0^t dt' \exp\left[-\gamma(t-t')\right]W_{ij}(t-t')\xi_j(t') ,
\end{gather}
where $W(t) = \exp(t\sigma)$. At large times the initial polymer
elongation is forgotten and after averaging over the thermal
fluctuations $\bm \xi$ one can easily obtain the following distribution
function:
\begin{eqnarray}\label{eq:PDFR}
 & P({\bm R}) = (2\pi)^{-3/2} (\det I)^{-1/2} \exp\left[-\frac{1}{2}{\bm R}^T I^{-1}
 {\bm R}\right] \\
 & I = \frac{2\gamma R_0^2}{3} \int_0^\infty \exp(-2\gamma t) W(t)W^T(t)\, .
\end{eqnarray}
The probability density function (PDF) \eqref{eq:PDFR} is valid for any velocity
gradient. In the particular case of a steady shear flow in the $XY$ plane, as shown
in Fig.~\ref{fig:orient}, the matrices can be found explicitly:
\begin{gather}
W(t) = \left(
\begin{array}{ccc}
1 & s t & 0 \\
0 & 1 & 0 \\
0 & 0 & 1
\end{array}\right),\quad
I = \frac{R_0^2}{3}\left(
\begin{array}{ccc}
1+s^2/2\gamma^2 & s/2\gamma & 0 \\
s/2\gamma & 1 & 0 \\
0 & 0 & 1
\end{array}\right)\, ,
\end{gather}
where the axis are sorted in the order $X,Y,Z$. One can see that for
large Weissenberg numbers $\mathrm{Wi} = s/\gamma \gg 1$, the
mean polymer elongation in the $X$ direction is much larger than
in transversal directions.

\subsection{Elongation PDF}
The PDF of elongation can be easily computed from \eqref{eq:PDFR}:
the elongation in the $y$ and $z$ direction are independent of
$\mathrm{Wi}$ and the two marginal PDFs are gaussian with zero mean
value and variance $\langle R_y^2 \rangle=\langle R_z^2
\rangle=\frac{R_0^2}{3}$, while in the mean flow
direction the variance is $\langle R_x^2 \rangle=\frac{R_0^2}{3}(1+\frac{1}{2}\mathrm{Wi}^2)$.\\
\begin{figure}
  \includegraphics[width=8cm]{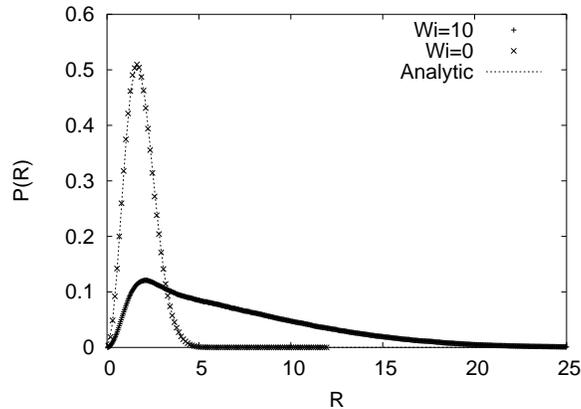}
\caption{The PDF of the modulus of the end-to-end vector $\bm R$ for
different $\mathrm{Wi}$ numbers. While for $\mathrm{Wi}=0$ the PDF is
concentrated around the radius of equilibrium, for higher values of
$\mathrm{Wi}$ it is broader and the high elongation configurations
become more and more probable.}
  \label{fig:pdfr}
\end{figure}
The distribution function of ${\bm R}$ cannot be obtained analytically
for an arbitrary value of $\mathrm{Wi}$, but one can study its
behavior in two limiting cases. For $\mathrm{Wi}\ll 1$ the effect of
the shear flow can be neglected, and we have the simple thermally
driven polymer with linear relaxation force. The distribution
function of its elongation has a simple gaussian form~\cite{book}:
\begin{equation}
 P(R) = \sqrt{\frac{2}{\pi}}\frac{R^2}{R_0^3}\exp\left[-\frac{R^2}{2 R_0^2}\right]\, .
\end{equation}
In the opposite case $\mathrm{Wi}\gg 1$ the system is strongly
anisotropic and the main contribution to the polymer elongation
comes from the $X$ component. This fact allows one to obtain the
right tail of the elongation PDF. For $R\gg R_0$ one has
\begin{equation}
 P(R) =
 \sqrt{\frac{12}{\pi(2+\mathrm{Wi}^2)}}\frac{1}{R_0}
 \exp\left[-\frac{3 R^2}{(2+\mathrm{Wi}^2)R_0^2}\right]\, .
\end{equation}
\subsection{Angular PDF}
\begin{figure}[t!]
  \subfigure[The analytic PDF is the numerical integration of \eqref{eq:angles}, while
the dots come from the numerical simulations.]{\label{fig:phi-a}\includegraphics[width=8cm]{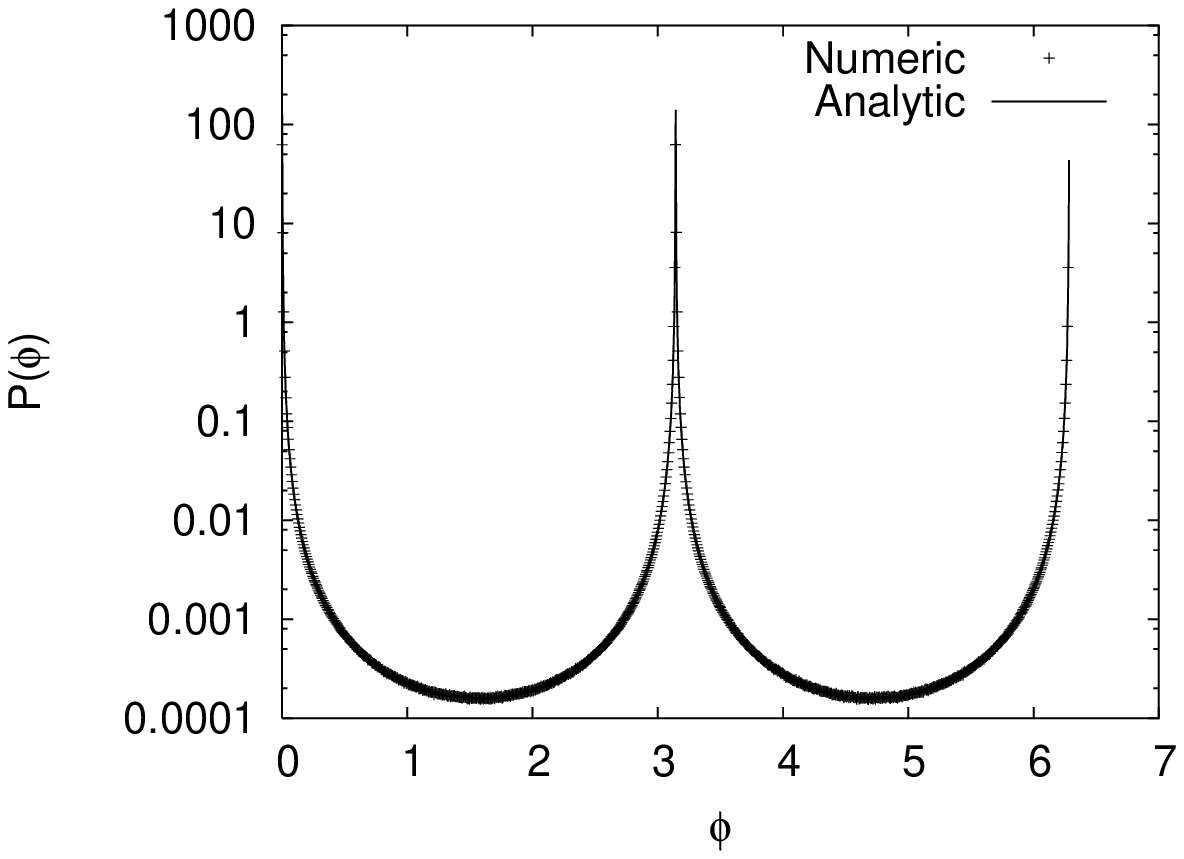}}
  \subfigure[The amplitude of the PDF's peak $\phi_t$ as a function of $\mathrm{Wi}$]{\label{fig:phi-b}\includegraphics[width=8cm]{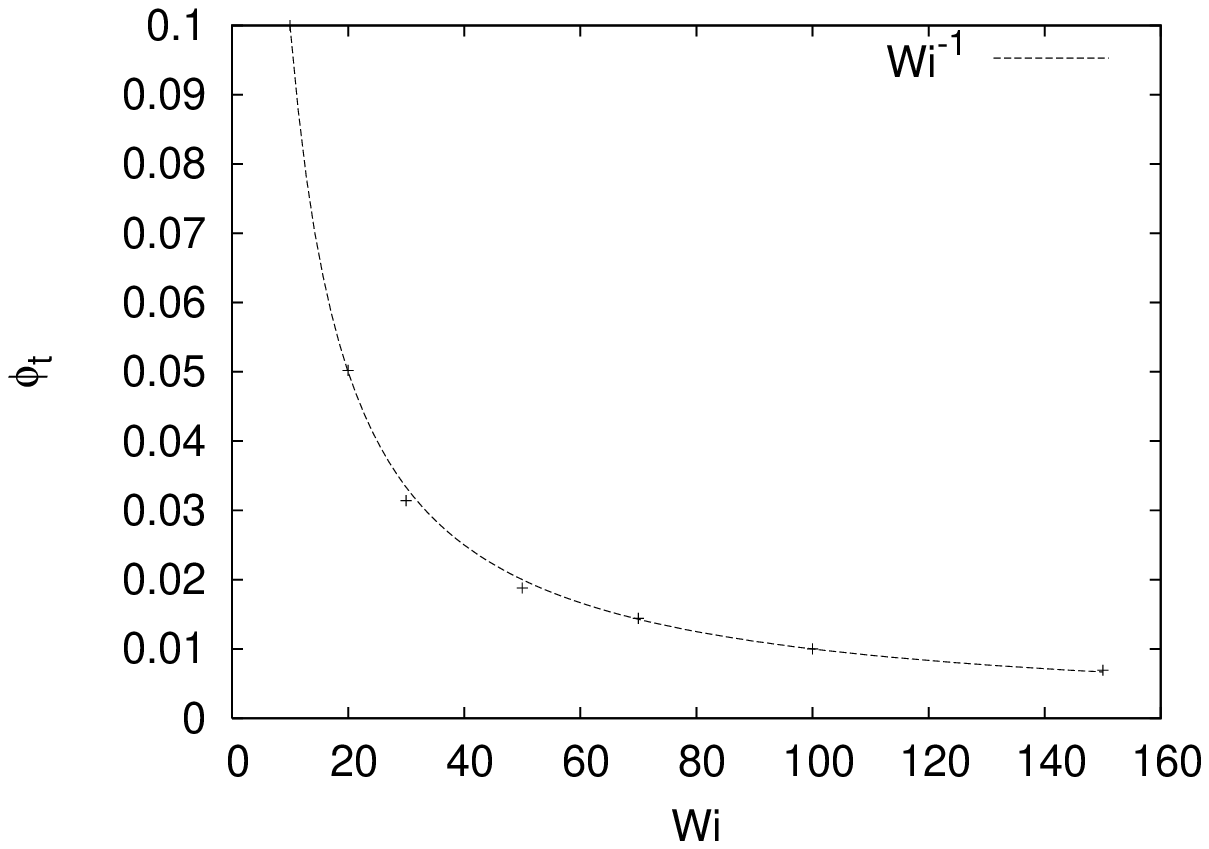}}
\caption{The PDF of the angle $\phi$}
\label{fig:phi}
\end{figure}

\begin{figure}[t!]
\includegraphics[width=8cm]{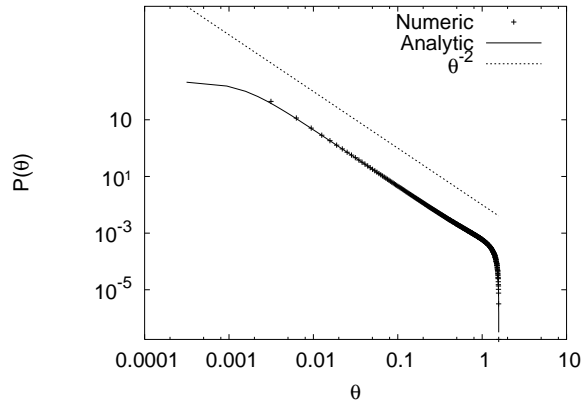}
\caption{The PDF of the angle $\theta$ is symmetric with respect to $\theta=0$ and can 
be described by a power law relationship over a wide range of values of $\theta$.}
  \label{fig:theta}
\end{figure}

Next we are interested in the orientation of the polymer. 
The angular distribution function will allow us to show that for
large $\mathrm{Wi}$ the polymer spends most of the time aligned
to the $X$ axis, while for $\mathrm{Wi}\ll 1$ the
orientation distribution is almost isotropic. In order to
parameterize the orientation of the polymer we introduce the
angles $\phi$ and $\theta$ as shown in Fig.~\ref{fig:orient}.
The angle $\phi$ represents the deviation of the
polymer end-to-end vector from the $X$ axis in the shear velocity
plane $XY$ while the angle $\theta$ gives us the polymer deviation
in the transversal direction $Z$. Switching to spherical coordinates we have
$R_x = R\cos\theta \cos\phi$, $R_y = R \cos\theta \sin \phi$, $R_z = R
\sin\theta$. After integrating over the polymer elongation in eq.\eqref{eq:PDFR} one
immediately obtains the angular PDF:
\begin{equation}
 {\mathcal{P}}(\phi,\theta)\propto\frac{
 \cos\theta}{\left\{1-\frac{\cos^2\theta}{4+\mathrm{Wi}^2}
 \left[\mathrm{Wi}^2\cos(2\phi)+2\mathrm{Wi}\sin(2\phi)\right]\right\}^{3/2}} \,.
 \label{eq:angles}
\end{equation}
The calculation of the two marginal PDFs is not possible in general, and we have to
integrate eq. \eqref{eq:angles} numerically
(Fig.~\ref{fig:phi-a},~\ref{fig:theta}).\\
As one can see in Fig.~\ref{fig:theta}, the PDF of $\theta$ decays as a power law 
(asymptotically ${\mathcal{P}}(\theta)\sim\theta^{-2}$ in the region $\phi_t \ll \theta
\ll 1$), and is symmetric about $\theta=0$. The higher the value of $\mathrm{Wi}$,
 the wider is the power law region.\\
In principle, if the polymer tumbles in the shear plane, we should
consider a time-dependent PDF of $\phi$ on a unbounded domain.
However if we consider $\phi$ to be between $0$ and $2\pi$ one
arrives at a stationary PDF, peaked in the regions
$\phi=k\pi+\phi_t, k=0,1,2$ (Fig.~\ref{fig:phi-a}), where
$\phi_t\sim\frac{1}{\mathrm{Wi}}$ as shown in Fig.~\ref{fig:phi-b}.
In the region where
$\phi\gg\phi_t$ we have ${\mathcal{ P}}(\phi)\sim\sin^{-2}{\phi}$.\\
The angular PDFs tell us that any fluctuations in $\theta$ are not
relevant to the tumbling dynamics, and that for the angle $\theta$
there is no relevant scale. The fact that increasing $\mathrm{Wi}$
the power law region becomes wider and wider means that when the shear
is much stronger than the relaxation the polymer is nearly aligned
with the $X$ axis. The PDF of the angle $\phi$ expresses the fact
that the polymer spends most of the time in the vicinity of
$\phi=0$, $\phi=\pi$, $\phi=2\pi$. The polymer is aligned in the $X$
direction as a result of the shear flow, it fluctuates for a certain time in the
region $\phi\sim\phi_t$, where the shape effects are strong. When
thermally activated it goes beyond the region $\phi \gtrsim \phi_t$ and 
the shear becomes more important and induces a fast rotation of
$\Delta \phi\simeq\pi$ or multiples thereof, that is the polymer tumbles.

\subsection{Tumbling time distributions}
\begin{figure}[t!]
  \subfigure[The elongation-based tumbling time PDF]{\epsfig{figure=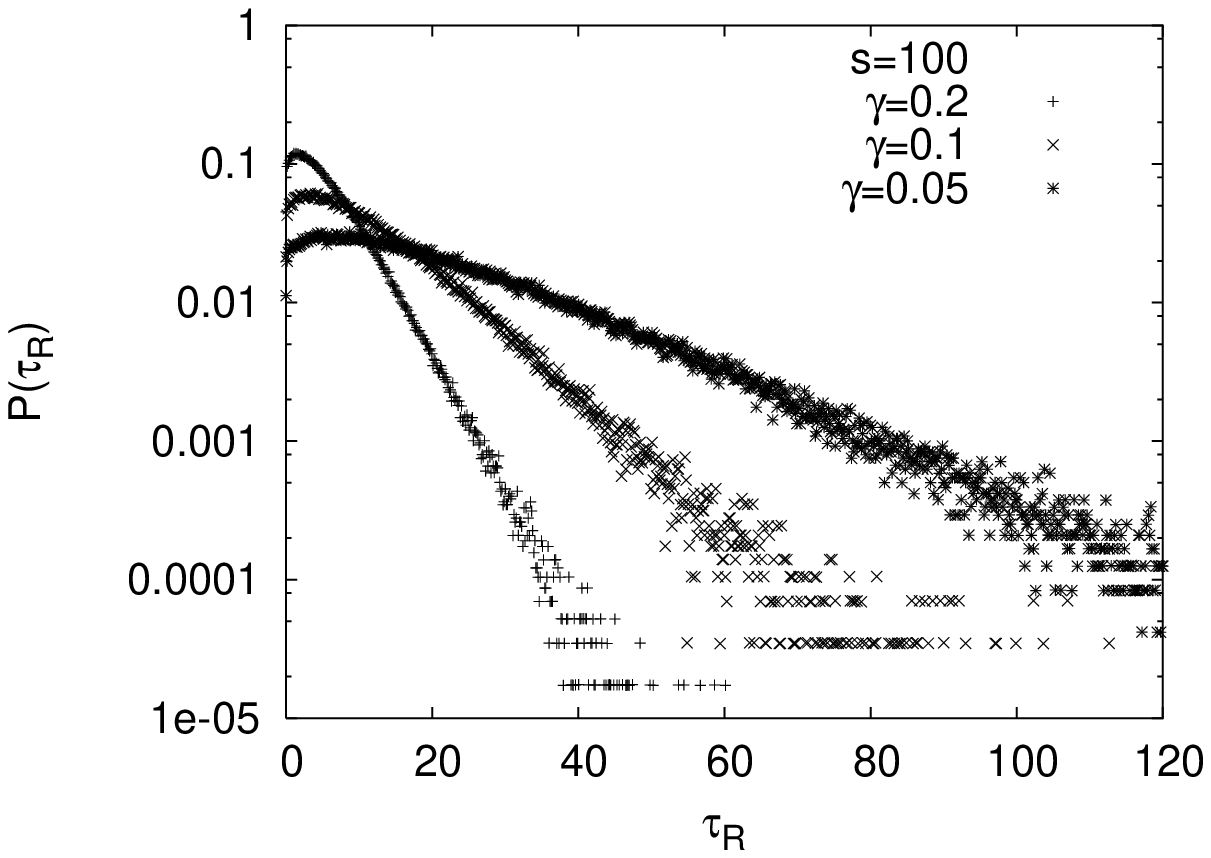,width=8cm}}
  \subfigure[The angle-based tumbling time PDF]{\epsfig{figure=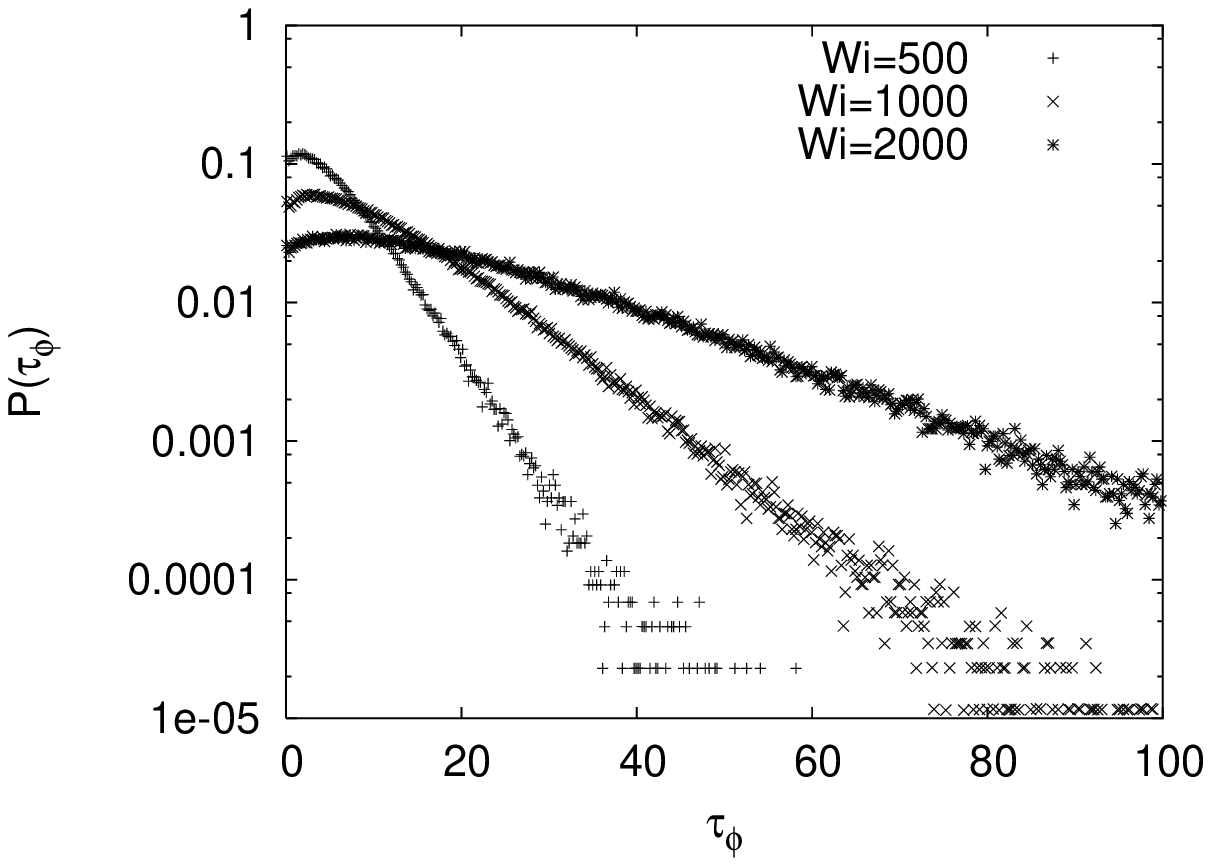,width=8cm}}
\caption{The two different tumbling time PDFs.}
\label{fig:thermtumble}
\end{figure}
In the previous two sections we have studied the stationary
distribution functions which can be measured experimentally by
averaging the polymer elongation and orientation over large time
periods. However, as described in the introduction and in~\cite{04CKLTa,T04}
the dynamics of the polymer is non stationary,
due to continuous tumbling. The natural question which arises is
whether there are some quantities, which would allow experimental
observation and quantitative description of the tumbling process. In
the recent papers by Chu et al.~\cite{99SBC,01HSBSC} the different
time correlation functions of the polymer elongation were studied,
and it has been shown that they can have different forms in the presence
of shear. Here we introduce the tumbling time, which can be measured 
experimentally~\cite{05GS} and used to describe the tumbling process.
As described above the most striking difference between the tumbling
dynamics of the polymer in a shear flow and spinning in a rotational
flow is the a-periodicity of the tumbling process. Due to the
stochastic nature of tumbling, the tumbling time, i.e. time
between subsequent flips of the polymer, is a random
variable with relatively large fluctuations. Our task is now to
study the distribution function of the tumbling time and to give
some theoretical and numerical predictions of its dependence on the
Weissenberg number.

In the case of the dumbbell model, we can define tumbling as a
flip of the polymer induced by the thermal noise. When the
polymer is in an unstable equilibrium configuration $\phi=k\pi,
k=0,1,2$ the thermal noise can bring the polymer out of the region
$\phi\sim\phi_t$, and this induces a fast rotation which takes a
time of the order of $s^{-1}$. We define the tumbling time
$\tau_\phi$ as the interval between two subsequent flips. This is
typically the time between subsequent crossings of the lines $\phi =
(k+1/2)\pi$. For $\mathrm{Wi}\gg 1$, as shown in the previous
section (see Fig.~\ref{fig:phi-a}), the polymer spends only a small fraction 
of the total time
far from the shear direction, and therefore the exact value of 
the angular threshold defining the tumbling time is not very important. Indeed
this time is made up of two contributions: the time spent in the
region $\phi\sim\phi_t$ and the duration of the rotation of order
$s^{-1}$. The latter can be neglected for large $\mathrm{Wi}$.

Experimental techniques do not always allow the polymer
orientation angle $\phi$ to be resolved and therefore it is convenient to
introduce another definition of the tumbling time ($\tau_R$). This
quantity measures the interval between subsequent
changes between stretched and coiled polymer conformations. In
other words, we start measuring when the length of the polymer
exceeds a certain threshold value and we stop
when it again becomes smaller than this threshold value.\\
While $\tau_\phi$ is the most natural definition it can be difficult
to measure experimentally. On the other hand, the PDF of $\tau_R$
depends in a non universal way on the threshold, but experimental
techniques allows it to be measured when the polymer is sufficiently
stretched.

In order to analyze the behavior of the tumbling time PDFs at large
Weissenberg numbers we should analyze the dynamics of the polymer
elongation projected onto the $X$-axis ($R_x$). Indeed, the angle-based
tumbling time $\tau_\phi$ measures the time intervals between
subsequent events when $R_x=0$, while the elongation based tumbling
time $\tau_R$ corresponds to the time intervals between crossings of
a threshold value $R_x = R_{TH}$. In order to find how the shape of
the tumbling time PDF changes with increasing $\mathrm{Wi}$ number, we can
rescale the time and the polymer size in the following way: $R_x = x
R_0 \mathrm{Wi}$, $R_y = y R_0$, $t = \gamma^{-1} \tau$, which leads
to the following equations:
\begin{gather}\label{eq:rescaling}
 \dot{x} = -x + y\\
 \dot{y} = -y + \sqrt\frac{2}{3}\zeta_y \\
 \langle\zeta_y(0)\zeta_y(\tau)\rangle = \delta(\tau) .
\end{gather}
In eq.~\eqref{eq:rescaling} we have omitted the term corresponding to $\xi_x$ 
because it is negligible in the large Weissenberg number regime compared to the other
terms. Note that these
equations do not depend on the Weissenberg number, and it enters the
problem only through the threshold $R_{TH}$. In our simulations we
choose $R_{TH} = \sqrt{3 \langle R^2\rangle } \propto R_0
\mathrm{Wi}$, so that in the dimensionless variables we have
$x_{TH} = 1/\sqrt{2}$ which does not depend on Weissenberg number either. 
From this analysis we conclude that the tumbling time PDF at
large Weissenberg numbers approaches some universal form: its peak 
is positioned at characteristic time scale of order $\gamma^{-1}$, 
and as $\gamma$ increases the peak moves towards the origin.
The problem of finding the exact form of this function is equivalent
to the problem of finding the PDF of persistence times of a
non-Markovian random process $x(\tau)$. This problem has recently
attracted a lot of attention (see e.g. ref.~\cite{pers}), however no
explicit solution is known for a general random process. Still it is
possible to make some predictions on the PDF: the right tail of the
PDF $\tau_R,\tau_\phi \gg \gamma^{-1}$ is exponential and has the
form $P(\tau_R) \propto \exp{[-c_{R} \gamma \tau_{R}]},\,
P(\tau_\phi) \propto \exp{[-c_{\phi} \gamma \tau_{\phi}]}$ because
large tumbling times correspond to a large number $\gamma \tau \gg
1$ of unsuccessful attempts to cross the threshold. The correlation
time of our stochastic process $R_x(t)$ is of order $\gamma^{-1}$,
and therefore these attempts to cross the threshold are almost
independent, and one should take the product of their probabilities,
which leads to the above exponential laws.
Both the PDFs of $\tau_\phi$ and $\tau_R$ are not well
defined for $\tau_\phi\sim 0$, $\tau_R\sim 0$ since they are sensitive 
to brownian noise discretization: it is possible to tumble very rapidly when
the polymer is coiled, and this is why the probability of measuring
a tumbling time much smaller than $s^{-1}$ is not zero. Both the
tumbling time definitions do not work very well in the case of small
$\mathrm{Wi}$ experimentally because of the high resolution needed,
and numerically  because of the discretization procedure.

\section{Strongly elongated and rigid molecules}\label{sec:rigid}
Another physical situation we are interested in is the dynamics of
strongly elongated polymers in random flows. In this model, described
in detail in~\cite{04CKLTa,04CKLTb,T04}, the polymer is
placed into a random flow above the coil-stretch transition, where the effect
of the thermal fluctuations can be neglected. In this case the orientational
dynamics of the polymer are decoupled from the evolution of the
elongation, so that we can introduce the unit vector $n_i =
R_i/R$, obeying the following evolution equation:
\begin{equation} \label{eq:direct}
 \dot{ n}_i  = n_j (\delta_{ik}-n_i n_k)\nabla_j v_k .
\end{equation}
The velocity gradient consists of a regular shear part (as in the
previous section), and an isotropic incompressible random part.
The polymer size in the experiments is always much smaller than the
characteristic viscous scale of the velocity field which allows us
to assume smoothness of the velocity field. We will also assume
short-correlated velocity field which corresponds to the
so called Kraichnan-Batchelor model, which has been extensively studied in recent years
\cite{00FGV}. In the framework of this model the velocity gradient
matrix $\sigma_{ij}=\nabla_j v_i$ in the Lagrangian frame is
described by a gaussian process with the following pair-correlation
function~\cite{K68}:
\begin{equation}\label{eq:sigmacorr}
 \langle \sigma_{ij}(t)\sigma_{kl}(t')\rangle= D
 \delta(t-t')(4\delta_{ik}\delta_{jl} -\delta_{il}\delta_{kj}-\delta_{ij}\delta_{kl})\, ,
\end{equation}
where $\sigma_{ij}$ is the gradient matrix of the random velocity
component. Using the same notations as in the previous sections one
arrives at the following dynamical equation for the angles:
\begin{eqnarray} &&
 \dot{\phi} = - s\sin^2\phi + \xi_\phi \,,
 \label{eq:phieq} \\ &&
 \dot{\theta} =- s
 \frac{\sin(2\phi)}{2}\sin\theta\cos\theta + \xi_\theta \,,
\label{eq:thetaeq}
\end{eqnarray}
where $\xi_\phi$ and $\xi_\theta$ are
zero mean random variables related to the fluctuating components of
the velocity gradient. The statistics of both $\xi_\phi$ and
$\xi_\theta$ can be obtained from the correlation function
\eqref{eq:sigmacorr}:
\begin{eqnarray}
 \langle\xi_\theta(t)\xi_\theta(t')\rangle = 4 D \delta(t-t') \\
 \langle\xi_\phi(t)\xi_\phi(t')\rangle = \frac{4 D}{\cos^2\theta} \delta(t-t')\, .
\label{phicorr}
\end{eqnarray}
Note that the measure of configurations with $\theta \sim \pi/2$ is small,
thus making the formal singularity in eq. \eqref{phicorr} not essential. 
From eq. \eqref{eq:phieq} it turns out that the polymer
experiences constant a-periodic tumbling in the $XY$ plane~\cite{04CKLTa,04CKLTb,T04}.

\subsection{Stationary angular PDF}
\begin{figure}[ht!]
\epsfig{figure=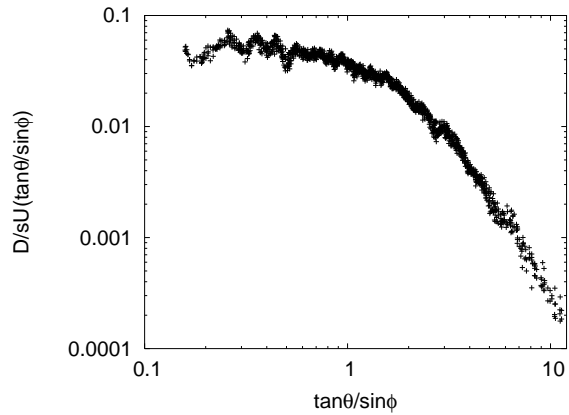,width=8cm}
\caption{The function $U(x)$, where $x=\tan\theta/\sin\phi$.}
\label{fig:uditheta}
\end{figure}
\begin{figure}[ht!]
   \subfigure[The agreement between the PDF and $\sin^{-2}{\phi}$ is good in the region
$\phi\gg \phi_t +k\pi, k=0,1,2$]{\epsfig{figure=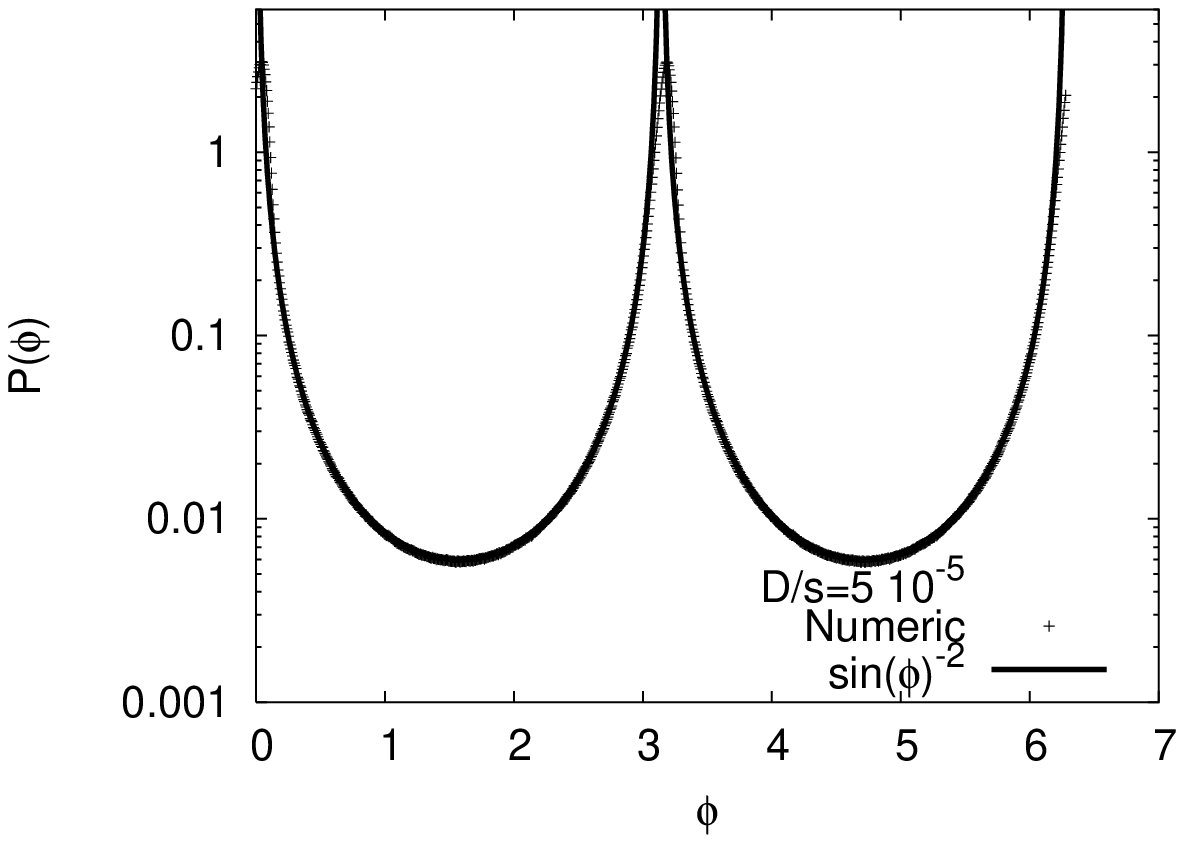,width=8cm}}
  \subfigure[The behavior of $\phi_t$ as a function of $D/s$, with 
the dotted line shown as a guide only.]{\epsfig{figure=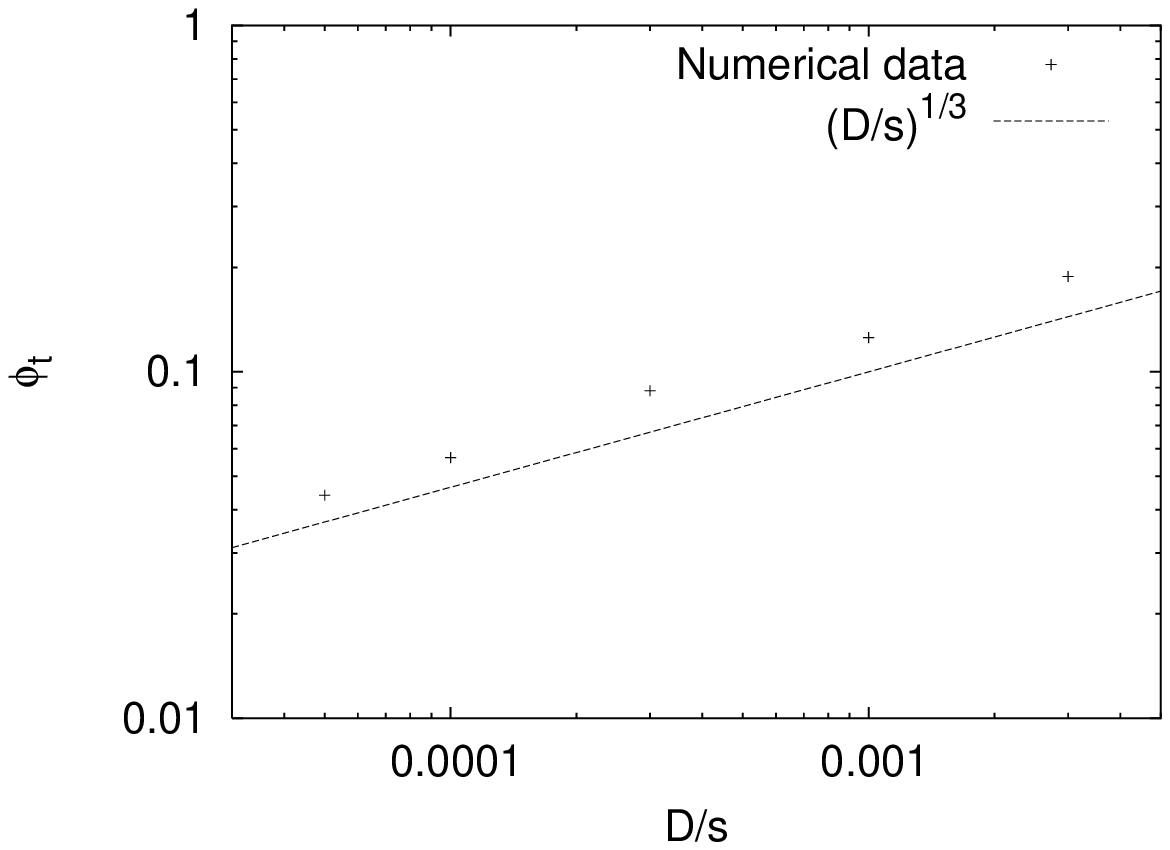,width=8cm}}
\caption{The PDF of the angle $\phi$}
\label{fig:fluctangles}
\end{figure}
From a dimensional analysis it can be shown that for $s\ll D$ the
characteristic values of the angles $\phi,\theta$ will be of order
$(D/s)^{1/3} \ll 1$. In this region one can set $\cos\theta = 1$ in
the expression \eqref{phicorr} so that eq. \eqref{eq:phieq} becomes
completely independent of $\theta$. Hence in this limit one derives:
\begin{equation}
 P_{st}(\varphi) =
\frac{\omega}{D}\int_0^\infty\mathrm{d}\phi
\exp\left[-\frac{s}{8D}\phi(\phi-2\varphi)^2-\frac{s\phi^3}{24
D}\right]\, ,
\end{equation}
where $\omega$ is the mean rotation frequency of the polymer,
which is determined from the normalization condition $\int_0^\pi
P_{st}(\phi)\mathrm{d}\phi=1$ and is given by
\begin{equation}
 \omega = \frac{(D s^2)^{1/3}}{4\cdot3^{1/6}\Gamma(7/6)\sqrt{\pi}}\, .
\end{equation}
The explicit form of the joint angular PDF is hard to compute
analytically. However one can obtain the expression for the tails of
the PDF $\phi,\theta\gg (D/s)^{1/3}$~\cite{04CKLTb,72HL}:
\begin{equation}
 P(\phi,\theta) = \frac{U(\tan\theta/\sin\phi)}{\sin^3\phi\cos^2\theta}\, ,
\end{equation}
where $U(x)$ is an unknown function with an universal argument. Numerical
simulations confirms this prediction, as shown in Fig.~\ref{fig:uditheta}.\\
\begin{figure}[t!]
\subfigure[The PDF of the angle $\theta$ is symmetric with respect to $\theta=0$ 
and can be expressed by a power law in a wide range of $\theta$.]{\label{fig:fluctheta}\includegraphics[width=8cm]{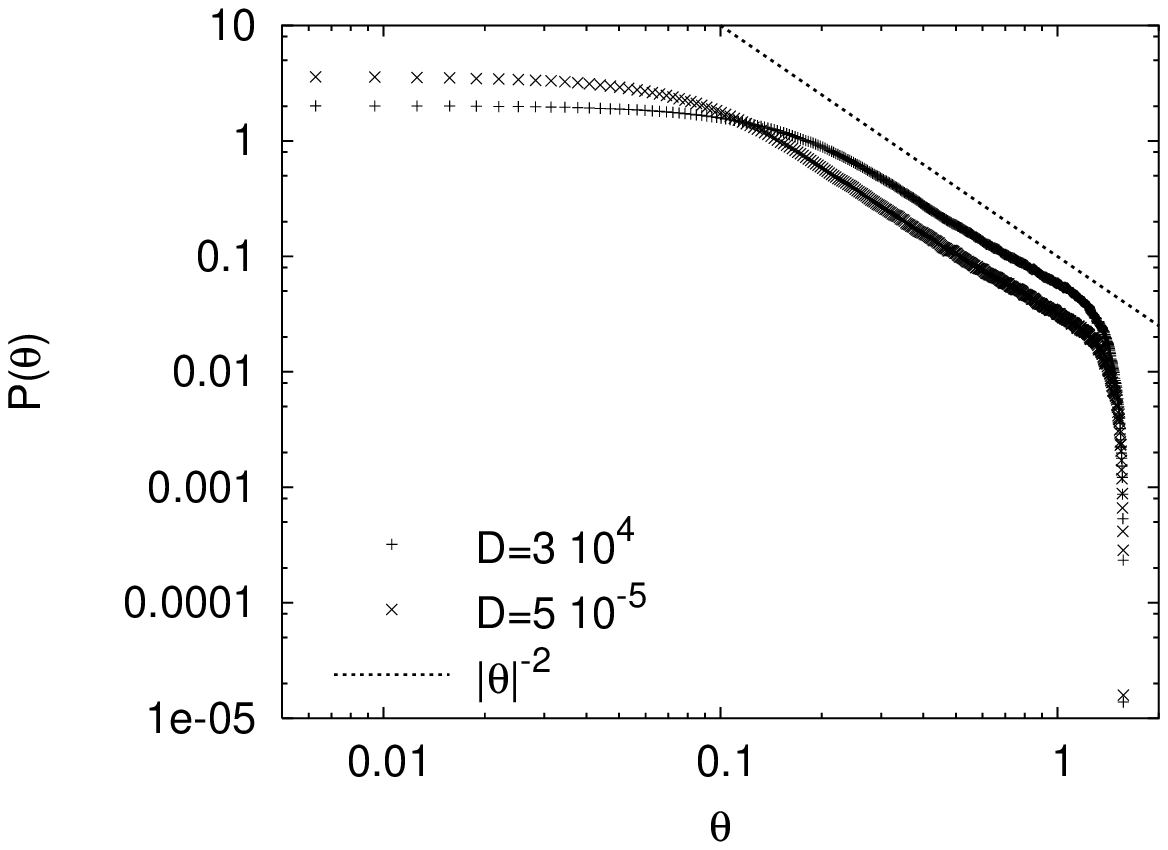}}
\subfigure[The power law for the core region $\phi\sim 0$. For this case $y\sim 3$.]{\label{fig:thetamx}\includegraphics[width=8cm]{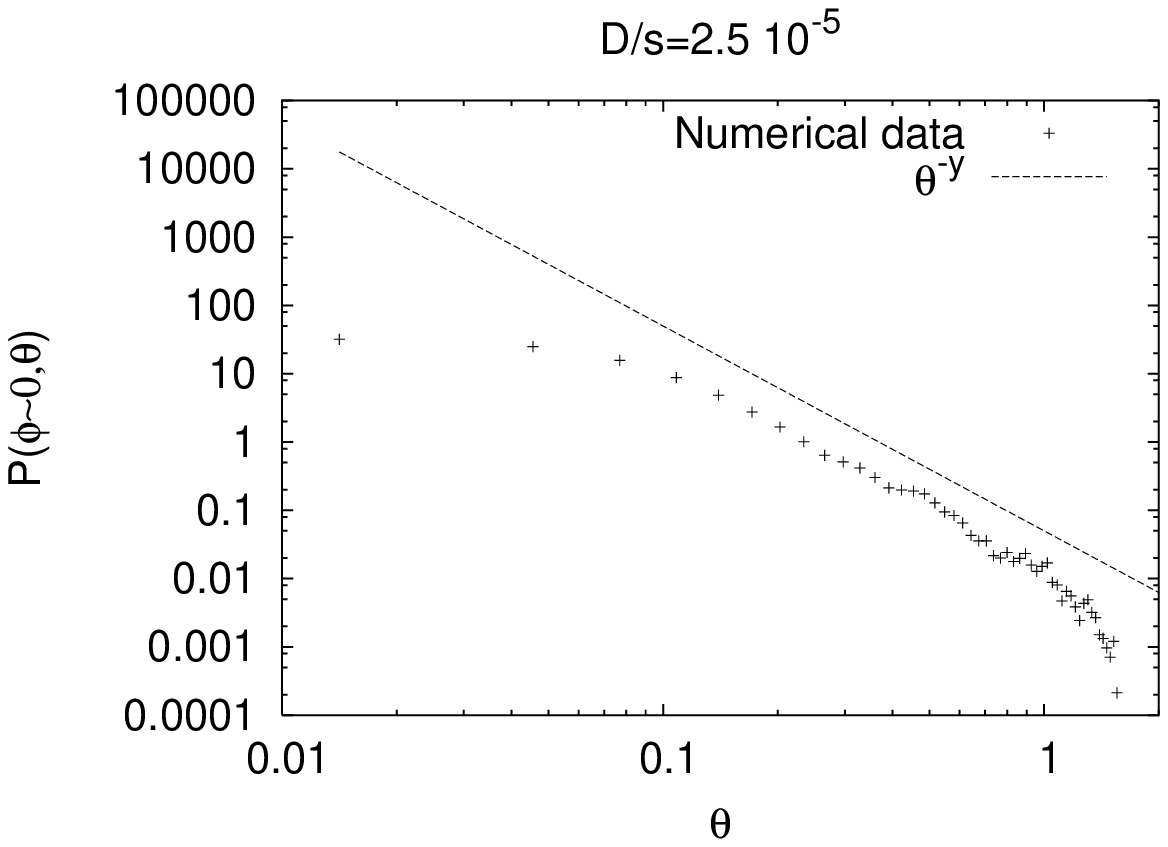}}
\caption{The PDF of the angle $\theta$.}
\label{fig:flucthetag}
\end{figure}
The two marginal PDFs obtained by numerical simulations show a behavior
similar to that of thermally driven polymers. Also the phenomenology of the
systems are similar, but now all the quantities depend on the
ratio between the amplitude of the random velocity gradient and the mean shear.
The PDF of the angle $\phi$ has peaks at $\phi\sim \phi_t+k\pi, k=0,1,2$, where
$\phi_t\sim (\frac{D}{s})^{1/3}$ and the PDF of $\theta$ has an algebraic core
${\mathcal{ P}}(\theta)\sim \theta^{-2}$ for $(\frac{D}{s})\ll 1$ 
(see Fig.~\ref{fig:fluctangles}).\\
In the stochastic region $|\phi| \lesssim \phi_t$ the tails of the
$\theta$-angle PDF $P(\theta)$ are also algebraic $P(\phi=0,\theta)
\propto \theta^{-y}$, but the exponent $y$ is non-universal and
depends on the statistical properties of the random velocity
gradient (see Fig.~\ref{fig:flucthetag}). It has been shown in~\cite{T04} that a simple
relation exists between the exponent $y$ and the entropy Cram\`er function
$S(x)$ of the Lyapunov exponent of the system (see sec.~\ref{sec:cramer}). For the
non-universal exponent $y$ the relation $y=S'(x)$ holds, where $x$ is the solution of the
equation $x S'(x) = S(x)$. Numerical simulations show that in the case
of an isotropic, short correlated, random velocity, the value of the
exponent is approximately $y \approx 3$, as shown in Fig.~\ref{fig:flucthetag}. This
implies that it is subdominant on the background of the $\theta^{-2}$ tail coming from
the regular dynamics region $|\phi|\gtrsim \phi_t$. However in
the general case of a finite-correlated and non-gaussian velocity field, 
one can imagine a situation where the non-universal exponent $y$ becomes smaller than the universal one
$y<2$, and in this case the $\theta$-angle distribution becomes non-universal.

\subsection{Tumbling time distribution}
\begin{figure}[b!]
   \subfigure[The exponential tail of the PDF.]{\epsfig{figure=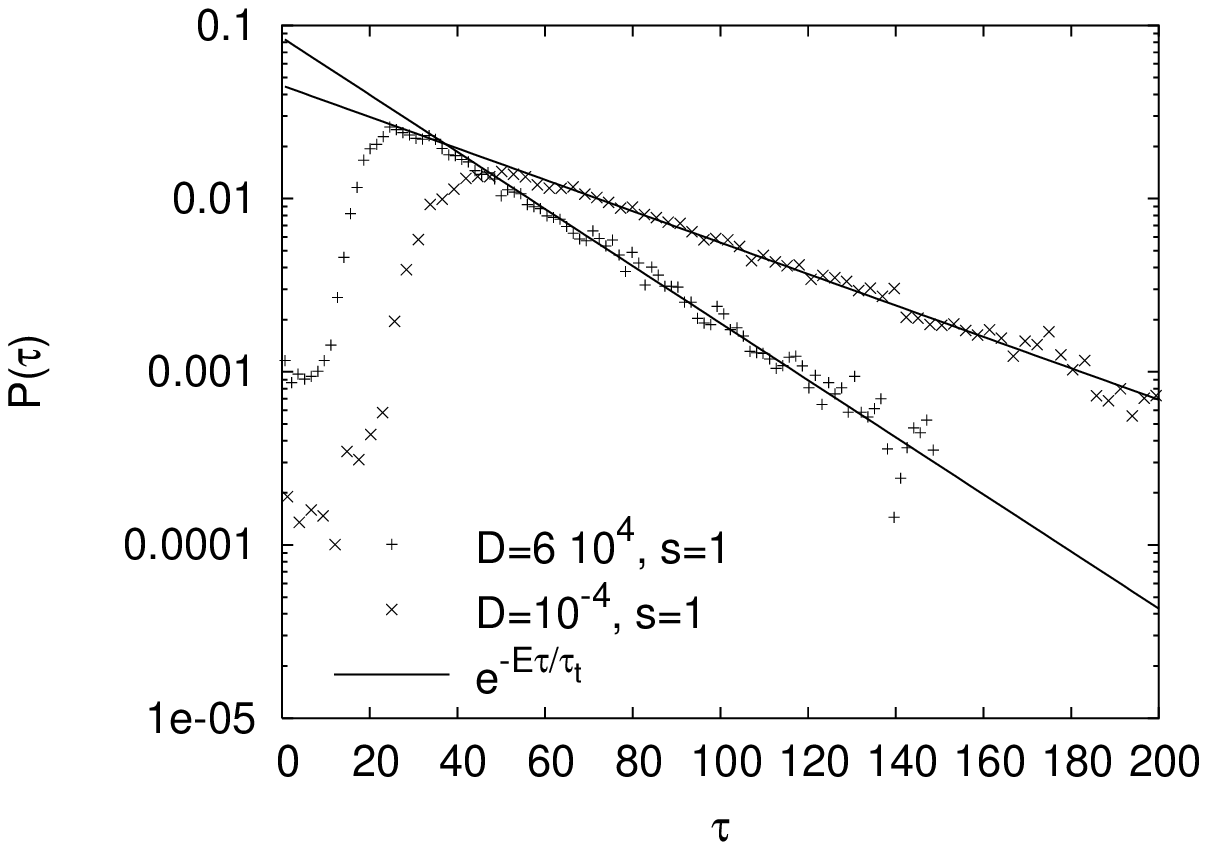,width=8cm}}
  \subfigure[The behavior of $\tau_t$ as a function of $s^2 D$]{\epsfig{figure=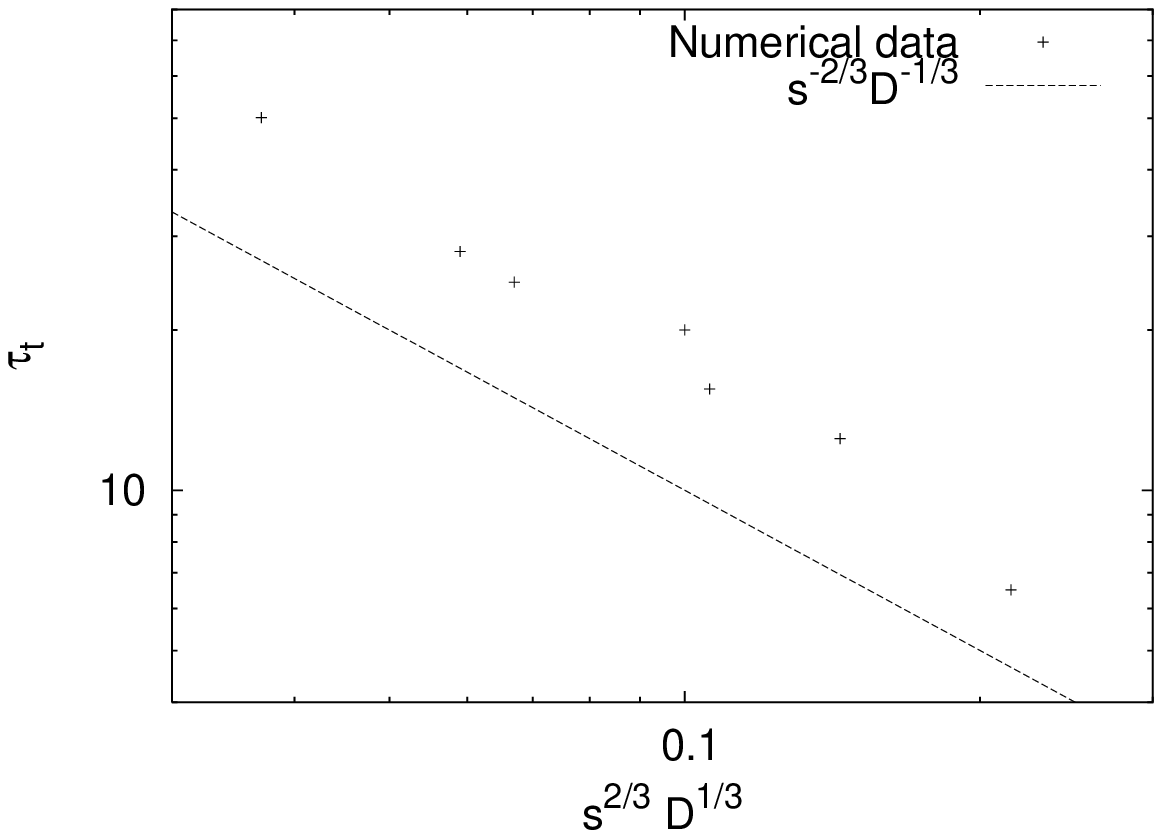,width=8cm}}
\caption{The PDF of the tumbling time $\tau$}
\label{fig:thermtau}
\end{figure}

For rigid polymers with a fixed size the tumbling time can be
defined only through the angular dynamics and therefore we will
refer only to $\tau_\phi$. The evolution time of this system is
$\tau_t\sim(s\phi_t)^{-1}$ so we expect that both the width and the
maximum of the tumbling time distribution are $\tau_t$. The tails
are related to the probability of passing (clock-wise) the angle
$\phi_t$ after a large amount of independent attempts, and can be
estimated as ${\mathcal{ P}}(\tau)\sim e^{-E (D s^2)^{1/3}
\tau}$~\cite{04CKLTa}. In~\cite{T04} it has been shown that the constant
$E$ is connected with the ground state energy value of a
one-dimensional quantum-mechanical system and can be estimated with
simple numerical analysis ($E\simeq0.45$). It is possible to
analyze the left tail of the tumbling time PDF:
\begin{equation}
  p(\tau) \propto \exp\left[-\frac{2 K^4(1/2)}{3 D s^2
  \tau^3}\right],\quad s^{-1}\ll\tau \ll (Ds^2)^{-1/3} ,
\end{equation}
where $K(x)$ is an elliptical integral of the first kind, and also to give
an exact expression for the tail $\tau \ll s^{-1}$ of the PDF.
For large values of $s/D$ this tail can be barely observed experimentally 
or numerically, and the structure of the tail strongly depends on the 
statistics of the chaotic flow, and is therefore non-universal 
(see Fig.~\ref{fig:thermtau}).
\section{Polymer elongation in chaotic flows} \label{sec:cramer}
\begin{figure}
  \includegraphics[width=8cm]{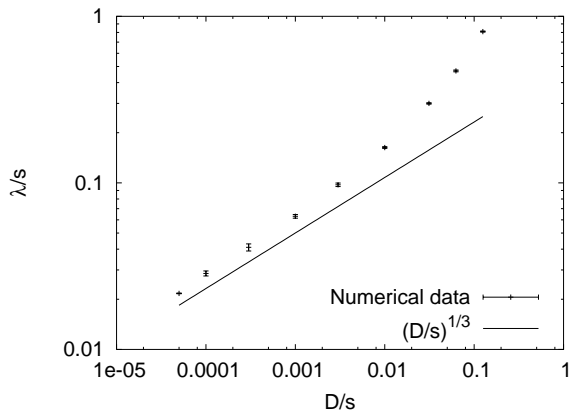}
  \caption{The behavior of the Lyapunov exponent as a function of $D/s$.}
\label{fig:lyapunov}
\end{figure}

\begin{figure}
\includegraphics[width=8cm]{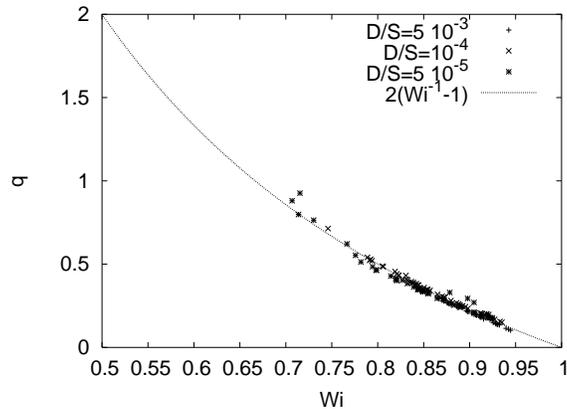}
  \caption{The exponent of the elongation PDF as a function of $\mathrm{Wi}$.}
\label{fig:expo}
\end{figure}

\begin{figure}[t!]
  \includegraphics[width=8cm]{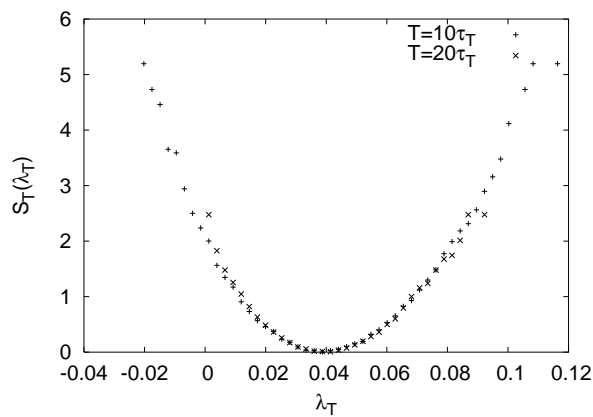}
  \caption{The Cram\`er function for two different times.}
\label{fig:pdgamma}
\end{figure}

In this final section we will study elongation statistics of the polymer
in the case of random velocity plus mean shear. The polymer is not
strongly elongated and can be treated as a linear dumbbell, as in
sec.~\ref{sec:therm}. Such a situation corresponds to a flexible
polymer in a chaotic flow below the coil-stretch transition\cite{74DG}.
Formally this is the case when the maximum Lyapunov exponent is
smaller than the inverse relaxation time $\lambda < \gamma$, where
the Lyapunov exponent is the rate of
divergence or convergence of two neighboring trajectories. \\
The equation governing the system is again eq. \eqref{eq:main} where
the gradient of the velocity is decomposed into a regular shear part and
a chaotic part. We can switch to spherical coordinates obtaining an
evolution equation for the angles (as in eqs.
\eqref{eq:phieq},\eqref{eq:thetaeq}) and an evolution equation for
the modulus of the elongation vector~\cite{04CKLTb,T04}:
\begin{equation}
\partial_t \ln{R}=-\gamma+\frac{s}{2}\cos^2{\theta}\sin{2\phi} +6D+\eta \, ,
  \label{eq:elongev}
\end{equation}
where in our model we assume $\langle \eta(t)\eta(t')
\rangle=2D\delta(t-t')$. Dimensional arguments show that
the Lyapunov exponent should be proportional to $(D s^2)^{1/3}$ 
(see Fig.~\ref{fig:lyapunov}).\\
In this situation the polymer spends most of the time in the coiled
state, but rare events lead to a large polymer stretching, when the
flow becomes strong enough. It can be shown, that the right tails of
polymer elongation PDF have the algebraic form $P(R) \propto
R^{-1-q} $~\cite{00BFL,T04}, where the exponent $q$ depends on the 
parameters of the system. While in the thermal noise case the
tails were Gaussian, here the probability of observing the polymer
stretched is strongly enhanced. The algebraic behavior of the tail
can be easily explained: the probability of having a local
stretching rate $\lambda > \gamma$ for a large time $t$ decays
exponentially with $t$, and during such events the polymer is
stretched by a factor which grows
exponentially with the time $t$.\\
It has been shown in~\cite{00BFL,T04} that the exponent $q$ is
related with the Cram\`er function of the Lyapunov exponent of the flow
~\cite{00FGV}. This function can be found explicitly only in few cases, hence the
main aim of this section is to present numerical measurements of the Cram\`er
function and of the dependence of the exponent $q$ on
the Weissenberg number $\mathrm{Wi}_\lambda=\lambda/\gamma$ below the 
coil-stretch transition (i.e. $\mathrm{Wi}_\lambda <1$).
As shown in~\cite{T04} the large deviation theory predicts that for large
averaging times $T \gg(Ds^2)^{-1/3}\equiv \tau_T$ the PDF of the Lyapunov exponent is:
\begin{equation}
 P(\lambda_T) \sim \exp\left(-\frac{T}{\tau_T} S(\lambda_T
\, \tau_T)\right), \quad T\gg \tau_T ,
\end{equation}
where the function $S(x)$ is the Cram\`er function.
Note that for large but finite values of $T/\tau_T$ the measured Cram\`er
function  $S_T(x) = - (\tau_T/T)\log P_T(x/\tau_T) $ depends on
the time $T$, but the difference between $S_T(x)$ and $S(x)$
is significant only in the region $|x| \gtrsim T/\tau_T \gg 1$.
Therefore in order to calculate the core of the Cram\`er
function we can use the finite time approximation. In our simulations we considered the
ratio $T/\tau_T$ from $10$ up to $280$.\\
In order to connect the Cram\`er function with the exponent $q$ in
the elongation PDF we need to use the Legendre transform (see refs.~\cite{00BFL,T04}
for details):
\begin{gather}
 S(x)- x S'(x) + \gamma\tau_T S'(x) =0 \\
 q =S'(x) .
\end{gather}
In~\cite{00BFL} and in~\cite{cmv05} the behavior of $q$
in the case of zero mean shear has been computed, and the calculations leads to:
\begin{equation}
  q=\frac{2}{\Delta}\left(\gamma-\lambda\right) \, ,
\label{eq:exponent}
\end{equation}
where $\Delta$ is the variance of the Lyapunov exponent distribution.
In the case of Kraichnan field without mean shear the ratio
$\Delta$ is proportional to $\lambda$ (see ref.~\cite{00FGV} for details), so that $q$
becomes proportional to $\mathrm{Wi}_\lambda^{-1}-1$. 
In~\cite{kes02} the case of shear turbulence
has been analyzed and the scaling is the same as in eq.\eqref{eq:exponent}. \\
The exponent $q$ extracted from numerical simulations is plotted in Fig.~\ref{fig:expo}.
The convergence of the Lyapunov exponent, given by the Cram\`er function, is shown in
 Fig.~\ref{fig:pdgamma}.

\section{Conclusions}
The tumbling phenomenon~\cite{89Liu,99SBC,04CKLTa,05GS}  has been studied in the 
framework of the linear dumbbell model, and some universal features of this 
motion are derived and numerically verified.\\ 
Three different examples of polymers in a linear, steady, plane, shear flow have been 
studied: i) a flexible polymer experiencing thermal noise, ii) a rigid polymer in a 
smooth random velocity gradient above the coil-stretch 
transition~\cite{74DG,00BFL,04GCS}), and iii) a flexible polymer in a smooth 
random velocity gradient below the coil-stretch transition.\\
In all three cases, the polymer tumbles aperiodically and the probability 
density function of the time between two successive tumbling events is exponential. 
While in the case of pure thermal noise, the typical tumbling time is determined 
only by the relaxation process, in the cases of random velocity gradient, the typical
rate of divergence of two initially close lagrangian trajectories is the most
important time scale.\\
The analysis of the statistics of orientation leads one to conclude that the 
phenomenology of these three situations is very similar: in the plane formed by 
the polymer end-to-end vector and the $Z$ axis there are no relevant scales and the 
polymer spends 
most of the time in the shear plane. In the shear plane, the dynamics is determined by 
the balance between the shear  and the thermal fluctuations. The majority of time
is spent nearly aligned to the velocity field. Aperiodically the shear induces a 
tumbling event.\\
The elongation dynamics is determined only by the balance between the relaxation and 
the stretching due to the shear flow in the thermal noise case. In the other two cases
stretching is determined by the presence of a positive Lyapunov exponent. In the case
below the coil-stretch transition the tail of the PDF can be determined by measuring
the statistics of the Lyapunov exponent.
 
We would like to thank A. Celani, M. Chertkov, I. Kolokolov and V.
Lebedev for many fruitful and inspiring discussions and S.
Geraschenko and V. Steinberg for many useful comments and interest
in our work. We are grateful to Chris Carr and Colm Connaughton who helped 
us to develop the final form 
of this paper. We acknowledge hospitality of CNLS at Los Alamos
National Laboratory where this work was partially done.
KT was supported by RFBR grant 04-02-16520a and INTAS grant Nr 04-83-2922.


\begin{thebibliography}{99}
\bibitem{97PSC}
 T. T. Perkins, D. E. Smith, and S. Chu,
{\em Single Polymer Dynamics in an Elongational Flow},
 Science {\bf 276}, 2016, (1997).

\bibitem{98SC}
 D. E. Smith and S. Chu,
 {\em Response of Flexible Polymers to a Sudden Elongational Flow},
 Science {\bf 281}, 1335 (1998).

\bibitem{99SBC}
 D. E. Smith, H. P. Babcock, and S. Chu,
 {\em Single polymer dynamics in steady shear flow},
 Science {\bf 283}, 1724, (1999).

\bibitem{00HSL} J. S. Hur, E. S. G. Shaqfeh, and R. G. Larson,
{\em Brownian dynamics simulations of single DNA molecules in shear flow},
J. Rheol., {\bf
44}, 713, (2000).

\bibitem{01HSBSC}
 J. S. Hur, E. Shaqfeh, H. P. Babcock, D. E. Smith, and S. Chu,
 {\em Dynamics of dilute and semidilute DNA solutions in the start-up of shear flow},
 J. Rheol, {\bf 45}, 421, (2001).

\bibitem{04GCS}
 S. Gerashchenko, C. Chevallard, and V. Steinberg,
 {\em Single polymer dynamics: coil-stretch transition in a random flow},
 Submitted to Nature (see also nlin.CD/0404045).

\bibitem{05GS}
S. Gerashchenko and V. Steinberg ,
 {\em Statistics of tumbling of a single polymer molecule in shear flow},
 Submitted to Phys. Rev. Lett. (see also nlin.CD/0503028).

\bibitem{chu2} R. E. Teixeira, H. P. Babcock, E. Shaqfeh and S. Chu,
{\em Shear thinning and tumbling dynamics of single polymers in the flow-gradient plane},
Macromolecules, {\bf 38}, 581-592, (2005)

\bibitem{05STSC} C. M. Schroeder, R. E. Teixeira, E. S. G. Shaqfeh and S. Chu,
{\em Dynamics of DNA in the flow-gradient plane of steady shear flow: Observations
and simulations},
Macromolecules, {\bf 38}, 1967-1978, (2005).

\bibitem{ladoux} B. Ladoux, J. Quivy, P. S. Doyle, G.Amouzni and J. Viovy,
{\em Direct imaging of singlemolecules: from dynamics of a single
DNA chain to the study of complex DNA-protein interactions, Science
Progress },
{\bf 84}(4), 267, (2001).

\bibitem{chu} S. Chu,
{\em Biology and polymer physics at the single-molecule level },
Phil. Trans. R. Soc. Lond. A, {\bf 361}, 689–698 6, (2003).

\bibitem{manneville} S. Manneville, P. Cluzel, J. L. Viovy, D. Chatenay and F. Caron,
{\em Evidence for the universal scaling behaviour of a freely relaxing DNA molecule},
Europhys. Lett., {\bf 36}, 413, (1996).

\bibitem{larson} R. G. Larson, T. T. Perkins, D. E. Smith and Chu, S.,
{\em Hydrodynamics of a DNA molecule in a flow field},
Phys. Rev. E, {\bf 55}, 1794, (1997).

\bibitem{cui} Y. Cui and C. Bustamante,
{\em Pulling a single chromatin fiber reveals the forces that maintain its
higher-order structure},
Proc. Natl. Acad. Sci., {\bf 97}, 127, (2000).

\bibitem{hegner} M. Hegner, S. B. Smith, and C. Bustamante,
{\em Polymerisation and mechanical properties of single RecA-DNA filaments},
Proc. Natl. Acad. Sci., {\bf 96}, 10109 (1999).

\bibitem{yin} H. Yin, M. D. Wang, K. Svoboda, R. Landick, S. M. Block and
J. Gelles,
{\em Transcription against an applied force},
Science, {\bf 270}, 1653 (1995).

\bibitem{wuite} G. J. Wuite, S. B. Smith, M. Young, D. Keller and  C. Bustamante,
{\em Single molecule studies of the effect of template tension on T7 DNA polymerase
activity},
Nature, {\bf 404}, 103 (2000).

\bibitem{04GS}
 A. Groisman and V. Steinberg,
{\em Elastic turbulence in curvilinear flows of polymer solutions},
 New J. Phys. {\bf 6}, 29 (2004).

\bibitem{72HL} E. J. Hinch and  L. G. Leal,
{\em The effect of Brownian motion on the rheological properties of a suspension
of non-spherical particles}, J. Fluid Mech., {\bf 52}, 634, (1972).

\bibitem{04CKLTa}
M. Chertkov, I. Kolokolov, V. Lebedev and K. Turitsyn,
{\em Tumbling of polymers in random flow with mean shear},
Submitted to J. Fluid Mech.

\bibitem{89Liu} T. W. Liu,
{\em Flexible polymer chain dynamics and rheological properties in steady flows},
J. Chem. Phys., {\bf 90}, 5826, (1989).

\bibitem{book}
 R. B. Bird , C. F. Curtiss , R. C. Armstrong  and O. Hassager,
{\em Dynamics of Polymeric Liquids}, Wiley, New York, 1987.

\bibitem{T04} K. Turitsyn,
{\em Polymer dynamics in chaotic flows with strong shear component},
Submitted to Phys. Rev. E (see also nlin.CD/0501025).

\bibitem{04CKLTb}
{\em Statistics of polymer extension in a random flow with mean shear},
M. Chertkov, I. Kolokolov, V. Lebedev and K. Turitsyn, submitted to J. Fluid Mech.

\bibitem{oett}
H. C. Ottinger,{\em Stochastic Processes in Polymeric Liquids}, Springer, Berlin,(1995).

\bibitem{future} A. Celani, A. Puliafito and K. Turitsyn,
{\em Polymers in linear shear flow: a numerical study},
EuroPhys. Lett., {\bf 70}, 464, (2005). 

\bibitem{80BGGS}
G. Benettin, L. Galgani, A. Giorgilli and J.-M. Strelcyn,
{\em Lyapunov exponents for smooth dynamical systems and for Hamiltonian systems;
a method for computing all of them},
Meccanica, {\bf 15}, 9-30, (1980).

\bibitem{pers}
S.N. Majumdar and C. Sire,
{\em Survival Probability of a Gaussian Non-Markovian Process: Application to the T = 0
Dynamics of the Ising Model},
Phys. Rev. Lett., {\bf 77}, 1420 (1996).

\bibitem{00FGV} G. Falkovich, C. Gawedzky and M. Vergassola,
{\em Particles and fields in fluid turbulence},
Rev. Mod. Phys. {\bf 73} (4), 913-975 (2001).

\bibitem{K68} R. H. Kraichnan,
{\em Small-scale structure of a scalar field convected by turbulence},
Phys. Fluids {\bf 11}, 945-963, (1968).

\bibitem{74DG}
P. G. De Gennes,
{\em Coil-stretch-transition of dilute felxible polymersunder ultra-high
velocity gradients},
J. Chem Phys, {\bf 60} 5030, (1974).

\bibitem{00BFL} E. Balkovsky, A. Fouxon and V. Lebedev,
 {\em Turbulent dynamics of polymer solutions},
 Phys. Rev. Lett. {\bf 84}, 4765 (2000);
 Phys. Rev. E {\bf 64}, 056301 (2001).

\bibitem{cmv05} A.Celani, S. Musacchio and D. Vincenzi,
{\em Polymer transport in random flow},
J. Stat. Phys., {\bf 118}, 529-552, (2005).

\bibitem{kes02} B.Eckhardt, J. Kronjager and J. Schumacher,
{\em Stretching of polymers in a turbulent environment},
Comp. Phys. Comm., {\bf 147}, 538, (2002).

\end{thebibliography}
\end{document}